\documentclass[12pt,a4paper]{scrartcl} 
\usepackage[T1]{fontenc}
\usepackage{ae,aecompl}
\usepackage{lmodern}
\usepackage{microtype}
\usepackage[mathcal]{eucal}
\usepackage[retainorgcmds]{IEEEtrantools}
\usepackage{graphicx}
\usepackage[usenames]{xcolor}
\usepackage[colorlinks,citecolor=blue,urlcolor=blue,linkcolor=blue]{hyperref}
\usepackage{amsmath,amsfonts,amssymb,amsbsy,tensor,mathrsfs,bm,dcolumn,wasysym}
\usepackage{indentfirst}
\usepackage[verbose,a4paper,top=3cm,left=3cm,bottom=2cm,right=2.5cm,headsep=5mm,footskip=1.5cm]{geometry}
\usepackage[hang]{footmisc} 
\usepackage{setspace}
\usepackage{siunitx}
\DeclareSIUnit\parsec{pc}

\newcommand{\lds}{\Lambda_{\text{dS}\pm}}
\newcommand{\slds}{\Lambda_{\text{dS}\pm}}
\newcommand{\ldsplus}{{\Lambda_{\text{dS}+}}}
\newcommand{\ldsminus}{{\Lambda_{\text{dS}-}}}
\newcommand{\rhol}{\rho_\Lambda}
\newcommand{\rholl}{\rho_{\tilde{\Lambda}}}

\begin{document}

\begin{titlepage}
\begin{center}
    {\LARGE \textbf{\textsf{Cosmology from a gauge induced gravity}}}

\bigskip\bigskip
    
    \href{mailto: ftovar@cbpf.br}{F.~T. Falciano}$^{\dagger}$, \href{mailto:gsadovski@if.uff.br}{G. Sadovski}$^{*}$, \href{mailto:sobreiro@if.uff.br}{R.~F. Sobreiro}$^{*}$, \href{mailto:tomaz@if.uff.br}{A.~A. Tomaz}$^{*}$ 

\bigskip 

 	$^\dagger$\footnotesize{\textit{CBPF - Centro Brasileiro de Pesquisas F\'isicas, \\ Rua Dr. Xavier Sigaud 150, 22290-180, \\ Rio de Janeiro, RJ, Brasil.}}
	
\bigskip
	
	$^*$\footnotesize{\textit{UFF - Universidade Federal Fluminense, \\ Instituto de F\'isica, Campus da Praia Vermelha, \\ Av. General Milton Tavares de Souza s/n, 24210-346, \\ Niter\'oi, RJ, Brasil.}}
\end{center}

\bigskip

\begin{abstract}
\noindent \textbf{\textsf{Abstract.}} The main goal of the present work is to analyze the cosmological scenario of the induced gravity theory developed in previous works. Such a theory consists on a Yang-Mills theory in a four-dimensional Euclidian spacetime with $SO(m,n)$ such that $m+n=5$ and $m\in\{0,1,2\}$ as its gauge group. This theory undergoes a dynamical gauge symmetry breaking via an In\"on\"u-Wigner contraction in its infrared sector. As a consequence, the $SO(m,n)$ algebra is deformed into a Lorentz algebra with the emergency of the local Lorentz symmetries and the gauge fields being identified with a vierbein and a spin connection. As a result, gravity is described as an effective Einstein-Cartan-like theory with ultraviolet correction terms and a propagating torsion field. We show that the cosmological model associated with this effective theory has three different regimes. In particular, the high curvature regime presents a de Sitter phase which tends towards a $\Lambda$CDM model. We argue that $SO(m,n)$ induced gravities are promising effective theories to describe the early phase of the universe.
\end{abstract}

\bigskip


\end{titlepage}

\section{Introduction}
\label{sec:Intro}

The Standard Model (SM) has shown to be a very efficient theoretical framework of the three fundamental interactions with the LHC continuously providing remarkable experimental confirmations. The SM is constructed over a consistent quantum gauge theoretical basis known as Yang-Mills theories. On the other hand, the fourth fundamental interaction lacks a consistent quantum description, i.e., quantum gravity remains an open issue with many theoretical proposals. Promising candidates for quantum gravity are (to name a few): Loop Quantum Gravity \cite{Ashtekar:2004eh,Ashtekar:2014kba}, Higher Derivative Quantum Gravity \cite{Buchbinder:1992rb,Asorey:1996hz,Shapiro:2008sf}, Causal Sets \cite{Henson:2006kf}, Causal Dynamical Triangulations \cite{Loll:2015yaa,Ambjorn:2012jv}, String Theory \cite{Duff:1996aw,Witten:1998qj,Polchinski:1998rq}, Asymptotic Safety \cite{Hawking:1979ig,Reuter:1996cp,Reuter:1994yq,Reuter:2001ag}, and Emergent Gravities \cite{Buchbinder:1992rb,Huang:2015zma,Barcelo:2001tb}. As widely known, all of these theories have their share of goals and problems which, for the sake of objectivity, we will not enumerate here. Nevertheless, string theories and emergent gravities share a common aspect that deserves attention: they do not deal with the direct quantization of the gravitational field. These theories have gravity as an emergent phenomenon. In other words, gravity, as a geometrodynamical phenomenon, is only a classical limit of a completely different quantum theory.

In the wide range of emergent gravities, there are several articles about gauge theories which employ a physical mechanisms to ``free'' the geometrical degrees of freedom, see for instance \cite{MacDowell:1977jt,Stelle:1979aj,Pagels:1983pq,Tresguerres:2008jf,Tseytlin:1981nu,Sobreiro:2007pn,Mielke:2011zza,Sobreiro:2011hb,Assimos:2013eua}. As a general aspect, the gravities which are generated in this way are described in first order variables \cite{Utiyama:1956sy,Sciama:1962,Kibble:1961ba}. These particular scenarios are very interesting in virtue of its resemblance with the SM, putting gravity on equal footing with the other three fundamental interactions. In this sense, geometrodynamics would be only a classical manifestation of another gauge theory, just like spontaneous symmetry breaking and hadronization are low energy effects of the different sectors of the SM.

The theory developed in \cite{Sobreiro:2011hb} is based on the Yang-Mills action for the group\footnote{The quantities $m$ and $n$ obey the bound  with $m+n=5$ with $m\in\{0,1,2\}$.} $SO(m,n)$. From this theory, a generalized gravity action emerges due to a non-perturbative dynamical symmetry breaking. Differently from the other approaches, the physical mechanism responsible for this breaking is not the Higgs mechanism, but the soft BRST breaking associated to the so-called Gribov problem. The running of the Gribov and coupling parameters work together for an In\"on\"u-Wigner contraction \cite{Inonu:1953sp} $SO(m,n)\longmapsto ISO(m!-1,p)$ where $p=5-m!$. Since the group $ISO(m!-1,p)$ does not represent a symmetry of the original action, the theory suffers a symmetry breaking for the common stability group, namely $SO(m!-1,p)$. This symmetry breaking is responsible for ``freeing'' the gravitational variables (see section \ref{sec:Inducedgrav} for more details). The resulting theory describes a geometrodynamical action for the vierbein and spin connection composed by the Einstein-Hilbert term, the cosmological constant, a quadratic torsion term, and a quadratic curvature term. 

In the present paper, we focus on the study of the cosmological scenario provided by the emergent gravity mentioned above. We are particular interested in the cosmological solutions that may describe the early universe. In fact, the current paradigm to describe the early universe is inflation
\cite{Guth:1980zm,Linde:1981mu,Starobinsky:1979ty,Bardeen:1983qw}. In this particular scenario, the standard cosmological model, namely, the $\Lambda$CDM model is preceded by an exponentially accelerated phase which should be responsible for solving the problems associated with the Friedmann-Lema\^itre-Robertson-Walker (FLRW) metric such as the flatness, isotropy, horizon, and monopole excess. In addition, one of the most attractive features of inflation is the prediction of primordial quantum fluctuations that seed the large scale structure with the correct scale-invariant spectrum. Apart from all its success, the inflationary scenario has some weakening points related to the existence of an initial singularity \cite{Borde:1993xh} and the open issue if inflation can indeed solve the homogeneity problem \cite{Calzetta:1992gv,Goldwirth:1991rj,Kofman:2002cj}. The proper manner to tackle these issues is to work with a complete quantum gravity theory.

Recently, there has been an increased interest in bouncing models in the literature. These models are valuable alternatives to inflation \cite{Finelli:2001sr,Allen:2004vz,Peter:2008qz}. Bouncing universes are non-singular models which can be generated within higher derivative and quantum gravity theories \cite{Brandenberger:1993ef,Mukhanov:1991zn,Biswas:2005qr,Bojowald:2001xe,Falciano:2015fja,PintoNeto:2012ug,Falciano:2007yf,PintoNeto:2005gx}. All these results suggest that the early universe is a promising arena to investigate theories associated to very high energy physics.

Our paper has the following structure: In section \ref{sec:Inducedgrav} we briefly point out the principal building blocks of our theory of gravity. In section \ref{sec:Cosmo} we construct a cosmological model from the induced gravity that we built. In section \ref{sec:Concl} we display our conclusions and further discussions.

\section{Induced gravity from a Yang-Mills theory}\label{sec:Inducedgrav}

In a previous paper \cite{Sobreiro:2012iv}, some of us have used a Yang-Mills theory in order to describe gravity at the quantum level. The main idea is to follow Quantum Chromodynamics (QCD) techniques such that, at low energies, a geometrodynamical theory of gravity is generated. In this section, we will briefly review the main features of these gauge theories. 

We start with a pure gauge theory based on the $SO(m,n)$ group with $m+n=5$ and $m\in\{0,1,2\}$. When $m=0$, we have the orthogonal group. For $m=1$ and $m=2$, we have, respectively, de Sitter and anti-de Sitter groups. The spacetime is an Euclidian four-dimensional differential manifold, $\mathbb{R}^4$. The algebra of the group is given by
\begin{equation}
\left[J^{AB},J^{CD}\right]=-\frac{1}{2}\Big[\left(\eta^{AC}J^{BD}+\eta^{BD}J^{AC}\right)-\left(\eta^{AD}J^{BC}+\eta^{BC}J^{AD}\right)\Big]\;,
\label{alg1}
\end{equation}
where $J^{AB}=-J^{BA}$ are the $10$ anti-hermitian generators of the gauge group. Capital Latin indexes are chosen to run as $\{5,0,1,2,3\}$. As a Lie group, $SO(m,n)$ can be seen as a five-dimensional flat space, $\mathbb{R}^{m,n}_S$, with invariant Killing metric given by $\eta^{AB}\equiv\mathrm{diag}(\epsilon,\varepsilon,1,1,1)$ where $\epsilon=(-1)^{(2-m)!}$ and $\varepsilon=(-1)^{m!+1}$. It is worth mentioning that the two spaces $\mathbb{R}^4$ and $\mathbb{R}^{m,n}_S$ are not dynamically related to each other, i.e., the gauge group has no relation with spacetime whatsoever. 

The $SO(m,n)$ group can be decomposed as a direct product, $SO(m,n)\equiv SO(m!-1,p)\times S(4)$ where $S(4)\equiv SO(m,n)/SO(m!-1,p)$ is a symmetric coset space and $p=5-m!$. This can be accomplished by projecting the group space into its fifth coordinate direction $A=5$. For convenience, let us label $J^{5a}=J^a$, where small Latin indexes vary as $\{0,1,2,3\}$. Thus, the algebra \eqref{alg1} can be written as 
\begin{IEEEeqnarray}{rCl}\IEEEyesnumber \label{alg2}
\left[J^{ab},J^{cd}\right]&=&-\frac{1}{2}\left[\left(\eta^{ac}J^{bd}+\eta^{bd}J^{ac}\right)-\left(\eta^{ad}J^{bc}+\eta^{bc}J^{ad}\right)\right]\;,\IEEEyessubnumber\\
\left[J^a,J^b\right]&=&-\frac{\epsilon}{2}J^{ab}\;, \IEEEyessubnumber\\
\left[J^{ab},J^c\right]&=&\frac{1}{2}\left(\eta^{ac}J^b-\eta^{bc}J^a\right)\;, \IEEEyessubnumber
\end{IEEEeqnarray}
with $\eta^{ab}\equiv\mathrm{diag}(\varepsilon,1,1,1)$.

Now, we construct the Yang-Mills action as
\begin{equation}
S_{\mathrm{YM}}=\frac{1}{2}\int\tensor{F}{^A_B}\ast\tensor{F}{_A^B}\;,\label{ym1}
\end{equation}
where $\tensor{F}{^A_B}$ is the field strength $2$-form, $F = \mathrm{d}Y+\kappa YY$, $\mathrm{d}$ is the exterior derivative, $\kappa$ is the coupling parameter and $Y$ is the gauge connection $1$-form. The gauge field is the fundamental field and it lies on the adjoint representation of the gauge group. The ``$\ast$'' operator denotes the Hodge dual operator in Euclidean spacetime. The action \eqref{ym1} is invariant under $SO(m,n)$ gauge transformations, $Y\longmapsto U^{-1}\left(\frac{1}{\kappa}d+Y\right)U$, with $U\in SO(m,n)$. The infinitesimal version of the gauge transformation is $Y\longmapsto Y+\nabla\alpha$ where $\nabla=\mathrm{d}+\kappa Y$ is the full covariant derivative and $\alpha$ is the infinitesimal gauge parameter.

Following the above prescription, the gauge field is decomposed as follows
\begin{equation}
Y=\tensor{Y}{^A_B}\tensor{J}{_A^B} = \tensor{A}{^a_b}\tensor{J}{_a^b}+\theta^aJ_a\;.\label{connec1}
\end{equation}
Thus, the decomposed field strength reads
\begin{equation} \label{Fdec}
F = \tensor{F}{^A_B}\tensor{J}{_A^B} = \left( \tensor{\Omega}{^a_b} - \frac{\epsilon\kappa}{4}\theta^a\theta_b\right)\tensor{J}{_a^b} + K^aJ_a\;,
\end{equation}
where we have defined $\tensor{\Omega}{^a_b}\equiv \mathrm{d}\tensor{A}{^a_b}+\kappa \tensor{A}{^a_c}\tensor{A}{^c_b}$ and $K^{a} \equiv \mathrm{d}\theta^{a}+\kappa \tensor{A}{^a_b}\theta^{b}$. It is straightforward to rewrite the Yang-Mills action \eqref{ym1} as
\begin{equation}\label{ym2}
 S_{\mathrm{YM}}=\frac{1}{2}\int \left[ \tensor{\Omega}{^a_b}\ast\tensor{\Omega}{_a^b} + \frac{1}{2}K^a\ast K_a - \frac{\epsilon\kappa}{2}\Omega_{\ b}^{a}\ast\left(\theta_{a}\theta^{b}\right)+\frac{\kappa^2}{16}\theta^{a}\theta_{b}\ast\left(\theta_{a}\theta^{b}\right)\right]\;.
\end{equation}
From the physical point of view, the actions \eqref{ym1} and \eqref{ym2} are indistinguishable - the action \eqref{ym2} is only a different way to write down the Yang-Mills action for the group $SO(m,n)$.

Before we go further, let us point out some interesting aspects of Yang-Mills theories and how we can make a consistent analogy with a possible quantum gravity theory. Yang-Mills theory presents two important properties, namely, perturbative renormalizability and asymptotic freedom \cite{Itzykson:1980rh}. Renormalizability is ensured by the BRST symmetry \cite{Piguet:1995er}. Asymptotic freedom \cite{Gross:1973id,Politzer:1973fx} means that the coupling parameter increases as the energy decreases. Hence, perturbation theory can only be employed at high energies. At low energies, the theory is settled in a highly non-perturbative regime. It is exactly at this regime that enters the Gribov problem \cite{Gribov:1977wm,Singer:1978dk}. As it is well-known, the Faddeev-Popov gauge fixing is not sufficient to eliminate all spurious degrees of freedom. In this sense, a residual gauge symmetry survives and it is quite relevant at low energies. The procedure to eliminate the Gribov copies is not completely understood but we do know that one needs a mass parameter and a soft BRST symmetry breaking\footnote{Recently, it was shown that, although standard BRST symmetry is broken, a non-pertubative generalization of the standard BRST symmetry can be defined \cite{Capri:2015ixa}.} related to the Gribov mass parameter, see \cite{Zwanziger:1992qr,Maggiore:1993wq,Dudal:2005na,Dudal:2011gd,Baulieu:2008fy,Baulieu:2009xr,Pereira:2013aza}. The BRST symmetry breaking, the Gribov parameter and the asymptotic freedom are the crucial effects that lead the action \eqref{ym2} to a gravity one. For details, see \cite{Sobreiro:2011hb,Assimos:2013eua}.

It is immediate to check that the $\theta$ field has the same degrees of freedom of a vierbein field, $e$. However, as a piece of the gauge field $Y$, the $\theta$ field has canonical dimension 1 while the vierbein field is dimensionless. Nevertheless, in despite of its transformation rule, the Gribov mass parameter can be used to adjust this dimension discrepancy. In fact, we can employ the following rescalings
\begin{equation}
A \rightarrow  \frac{1}{\kappa}A \;, \quad \theta \rightarrow  \frac{\gamma}{\kappa}\theta\;,
\label{rescale}
\end{equation}
which modify the de Sitter algebra \eqref{alg2} to
\begin{IEEEeqnarray}{rCl} \IEEEyesnumber \label{alg3}
\left[J^{ab},J^{cd}\right]&=&-\frac{1}{2}\left[\left(\eta^{ac}J^{bd}+\eta^{bd}J^{ac}\right)-\left(\eta^{ac}J^{bc}+\eta^{bc}J^{ad}\right)\right]\;,\IEEEyessubnumber\\
\left[J^a,J^b\right]&=&-\frac{\epsilon \gamma^2}{2\kappa^2}J^{ab}\;,\IEEEyessubnumber\\
\left[J^{ab},J^c\right]&=&\frac{1}{2}\left(\eta^{ac}J^b-\eta^{bc}J^a\right)\;. \IEEEyessubnumber
\end{IEEEeqnarray}
In addition, the action \eqref{ym2} can be recasted as 
\begin{equation}
S=\frac{1}{2\kappa^2}\int\left[\tensor{\bar{\Omega}}{^a_b} \ast \tensor{\bar{\Omega}}{_a^b}+\frac{\gamma^2}{2}\bar{K}^a \ast\bar{K}_a-\frac{\gamma^2}{2}\tensor{\bar{\Omega}}{^a_b} \ast (\theta_a\theta^b)+\frac{\gamma^4}{16}\theta^a\theta_b \ast(\theta_a\theta^b)\right]\;,\label{ym3}
\end{equation}
where $\tensor{\bar{\Omega}}{^a_b}\equiv \mathrm{d} \tensor{A}{^a_b}+ \tensor{A}{^a_c} \tensor{A}{^c_b}$, $\bar{K}^a\equiv\mathrm{D}\theta^a$ and $\mathrm{D}=\mathrm{d}+A$ is the covariant derivative with respect to the $SO(m!-1,p)$ sector.

The connection of the action \eqref{ym3} with a gravity theory is attained from the analysis of running of the ratio $\gamma^2/\kappa^2$. It has been shown in \cite{Assimos:2013eua} that this ratio vanishes at an energy scale near Planck energy. Hence, from \eqref{alg3}, it is clear that an In\"on\"u-Wigner contraction takes place, i.e. $SO(m,n)\longrightarrow ISO(m!-1,p)$. However, since the action \eqref{ym3} is not invariant under $ISO(m!-1,p)$ gauge transformations, this contraction induces a symmetry breaking in the theory that goes to the common stability group $SO(m!-1,p)$.

Given the symmetry breaking, the theory is now highly non-perturbative and it has also reached its quantum boundary to become a classical theory. This motivate us to find the possible physical observables. In QCD these are hadrons and glueballs. In a classical theory of gravity, these are geometrical entities. Hence, one may identify the effective metricity and affinity of spacetime with the gauge invariant quantities $g_{\mu\nu}=\eta_{ab}\langle\theta^a_\mu\theta^b_\nu\rangle$ and $\Gamma^\alpha_{\mu\nu}=\langle\theta^\alpha_a(\partial_\mu\theta_\nu^a+A^a_{\mu b}\theta^b_\nu)\rangle$, respectively. This idea allows a map between the gauge field pieces and the first order gravitational variables
\begin{IEEEeqnarray}{rCl}
\tensor{\delta}{^{\mathfrak{a}}_a}\tensor{\delta}{^b_{\mathfrak{b}}}\tensor{A}{^a_b} &=&\tensor{\omega}{^{\mathfrak{a}}_{\mathfrak{b}}} \;, \nonumber\\
\tensor{\delta}{^{\mathfrak{a}}_a}\theta^a &=& e^\mathfrak{a} \;,
\end{IEEEeqnarray}
where indexes $\{\mathfrak{a,b,c,\ldots}\}$ are related to the tangent space of the deformed spacetime, $\tensor{\omega}{^{\mathfrak{a}}_{\mathfrak{b}}}=\tensor{\omega}{^{\mathfrak{a}}_{\mathfrak{b}\mu}}dx^\mu$ is the spin connection 1-form and $e^\mathfrak{a}=\tensor{e}{^{\mathfrak{a}}_\mu}dx^\mu$ the vierbein 1-form. Additionally, it is convenient to identify
\begin{equation}\label{eq:newton-cosmol}
\gamma^2=\frac{\kappa^2}{4\pi G}=\frac{4\Lambda}{3}\;,
\end{equation}
where $G$ and $\Lambda$ are, respectively, Newton's constant and the gravitational cosmological constant. With these redefinitions, the action \eqref{ym3} finally becomes a gravity action
\begin{equation}\label{ym-map-grav}
 S_{\mathrm{Grav}}=\frac{1}{16\pi G}\int \left(\frac{3}{2\Lambda}\tensor{R}{^{\mathfrak{a}}_{\mathfrak{b}}}\star\tensor{R}{_{\mathfrak{a}}^{\mathfrak{b}}} + T^\mathfrak{a}\star T_\mathfrak{a}-\frac{\epsilon}{2}\varepsilon_\mathfrak{abcd} R^\mathfrak{ab} e^\mathfrak{c}e^\mathfrak{d} + \frac{\Lambda}{12}\varepsilon_\mathfrak{abcd}e^\mathfrak{a}e^\mathfrak{b}e^\mathfrak{c}e^\mathfrak{d}\right)\;,
\end{equation}
where $\tensor{R}{^{\mathfrak{a}}_{\mathfrak{b}}}=\mathrm{d}\tensor{\omega}{^{\mathfrak{a}}_{\mathfrak{b}}}+\tensor{\omega}{^{\mathfrak{a}}_{\mathfrak{c}}} \tensor{\omega}{^{\mathfrak{c}}_{\mathfrak{b}}}=\frac12 \tensor{R}{^\alpha_{\beta\mu\nu}}\tensor{e}{^{\mathfrak{a}}_\alpha}\tensor{e}{_{\mathfrak{b}}^\beta}dx^\mu dx^\nu$ is the curvature 2-form and $T^\mathfrak{a}=\mathrm{d}e^\mathfrak{a}+\tensor{\omega}{^{\mathfrak{a}}_{\mathfrak{b}}}e^\mathfrak{b}=\frac12\tensor{T}{^\alpha_{\mu\nu}}\tensor{e}{^{\mathfrak{a}}_\alpha}dx^\mu dx^\nu$ is the torsion 2-form\footnote{It is always important to stress that $\tensor{R}{^\alpha_{\beta\mu\nu}}$ is not the Riemann tensor, but it is the curvature of the spin connection.}. In the above equation, the ``$\star$'' operator represents the Hodge dual operator in the deformed spacetime $\mathbb{M}^4$.

Needless to say, the last two terms in \eqref{ym-map-grav} are recognized as the Einstein-Hilbert and the cosmological constant terms in the first order formalism. The other terms account for generalized terms of our effective gravity action, namely, a Yang-Mills-like term for the curvature and torsion. Perhaps, for the sake of clarity, it is worth expressing \eqref{ym-map-grav} in spacetime coordinates,
\begin{equation}\label{ym-map-grav-coord}
 S_{\mathrm{Grav}}=\frac{1}{16\pi G}\int\sqrt{-g}\;d^4x \left(\frac{3}{4\Lambda}\tensor{R}{_{\alpha\beta\mu\nu}}\tensor{R}{^{\alpha\beta\mu\nu}}+\frac{1}{2}\tensor{T}{_{\alpha\mu\nu}}\tensor{T}{^{\alpha\mu\nu}}-\epsilon R+2\Lambda\right)\;,
\end{equation}
where $R=g^{\mu\nu}R_{\mu\nu}$ is the curvature scalar and $g$ is the determinant of the metric $g_{\mu\nu}=\tensor{\eta}{_{\mathfrak{a}\mathfrak{b}}}\tensor{e}{^{\mathfrak{^a}}_\mu}\tensor{e}{^{\mathfrak{^b}}_\nu}$. However, bear in mind that we are still in the first order formalism: the metric is a composite field, the action \eqref{ym-map-grav-coord} does not define a high-derivative gravity and the curvature squared term, $\tensor{R}{_{\alpha\beta\mu\nu}}\tensor{R}{^{\alpha\beta\mu\nu}}$, does not equal Kretschmann scalar\footnote{Even though we can certainly decompose it as a sum of the torsion-free part plus contorsional contributions.}.

It is worth emphasizing that, in this theory, Newton's constant and the cosmological constant are coupled via \eqref{eq:newton-cosmol}. One-loop semi-perturbative estimations \cite{Assimos:2013eua} predict an exact agreement for Newton's constant but an extremely large value for $\Lambda$. This prediction seems to fail when compared with the observed cosmological constant (which we call $\tilde{\Lambda}$) \cite{Padmanabhan:2012gv,Perivolaropoulos:2008pg}. However, the Quantum Field Theory (QFT) prediction for the SM vacuum also predicts a discrepant value for the cosmological constant (here denoted by $\Lambda_{\mathrm{QFT}}$) \cite{Nelson:1982kt,Toms:1983qr,Buchbinder:1986gj,Parker:1985kc}. In order to adjust the theoretical predictions with the observational value, following \cite{Shapiro:2006qx,Shapiro:2009dh,Sobreiro:2011hb}, we assume that the value of the observed cosmological constant is due to a cancellation of the bare cosmological constant with the contributions coming from the expectation vacuum of the matter quantum fields, hence, we write the net renormalized cosmological constant as $\tilde{\Lambda}=\Lambda+\Lambda_{\mathrm{QFT}}$.

The vacuum field equations of this gravity theory can be found by applying the variational principle to action \eqref{ym-map-grav}. Its extremization with respect to the vierbein field yields to
\begin{equation}\label{eq:eq_mov1}
\frac{3}{2\Lambda}R^{\mathfrak{bc}} \star (R_{\mathfrak{bc}}e_{\mathfrak{a}})
 + T^{\mathfrak{b}}\star\left(T_{\mathfrak{b}}e_{\mathfrak{a}}\right)+\mathrm{D}\star T_{\mathfrak{a}}-\varepsilon_{\mathfrak{abcd}}\left(\epsilon R^{\mathfrak{bc}}e^{\mathfrak{d}}-\frac{\tilde{\Lambda}}{3}e^{\mathfrak{b}}e^{\mathfrak{c}}e^{\mathfrak{d}}\right) = 0 \;,
\end{equation}
while the extremization with respect to the spin connection field gives
\begin{equation} \label{eq:eq_mov2}
3 \mathrm{D} \star \left(\Lambda^{-1}\tensor{R}{_{\mathfrak{a}\mathfrak{b}}}\right) 
+e_{\mathfrak{b}} \star T_{\mathfrak{a}}-e_{\mathfrak{a}} \star T_{\mathfrak{b}}
 -\epsilon \varepsilon_{\mathfrak{abcd}} T^{\mathfrak{c}} e^{\mathfrak{d}} = 0 \;.
\end{equation}
It will soon be useful to express field equations \eqref{eq:eq_mov1} and \eqref{eq:eq_mov2} in spacetime coordinates. They are, respectively,
\begin{IEEEeqnarray}{rCl}\label{eq:mov1_coord}
\frac{3}{4\Lambda}\tensor{\Upsilon}{^\mu_\nu}+\frac{1}{2}\tensor{X}{^\mu_\nu}+D_\alpha \tensor{T}{_\nu^{\mu\alpha}}+2(\epsilon\tensor{G}{^\mu_\nu}+\tilde{\Lambda}\tensor{\delta}{^\mu_\nu})&=&0\; ,\\
3D_\beta(\Lambda^{-1}\tensor{R}{_{\mu\nu}^{\alpha\beta}})-\tensor{T}{_{\mu\nu}^\alpha}+\tensor{T}{_{\nu\mu}^\alpha}-\epsilon(\tensor{T}{^\alpha_{\mu\nu}}+\tensor{T}{^\beta_{\beta\mu}}\tensor{\delta}{_\nu^\alpha}-\tensor{T}{^\beta_{\beta}_\nu}\tensor{\delta}{_\mu^\alpha})&=&0\;, \label{eq:mov2_coord}
\end{IEEEeqnarray}
where we made use of the convenient definitions: $\tensor{{G}}{^\mu_\nu}\equiv\tensor{R}{^\mu_\nu}-\frac12 R\tensor{\delta}{^\mu_\nu}$ as the asymmetric Einstein tensor, $\tensor{\Upsilon}{^\mu_\nu}\equiv\tensor{R}{^{\alpha\beta\lambda\sigma}}\tensor{R}{_{\alpha\beta\lambda\sigma}}\tensor{\delta}{^\mu_\nu}-2\tensor{R}{^{\alpha\beta\lambda\mu}}\tensor{R}{_{\alpha\beta\lambda\nu}}$ and $\tensor{X}{^\mu_\nu}\equiv T_{\alpha\beta\lambda}T^{\alpha\beta\lambda}\tensor{\delta}{^\mu_\nu}-2\tensor{T}{_\alpha_\beta^\mu}\tensor{T}{^\alpha^\beta_\nu}$ as the tensors that give quadratic contributions in the curvature and torsion, respectively. 

First, we notice that the first term in both field equations can be suppressed by the enormous value of the bare cosmological constant $\Lambda$. In fact, for a weak curvature regime, these correction terms can be neglected and for a torsionless situation we arrive at the standard Einstein-Hilbert theory with a effective cosmological constant $\tilde{\Lambda}$. In fact, by looking at \eqref{eq:mov2_coord} at a weak curvature regime, the solution $T\approx0$ is immediate (at least for vanishing spin-densities). Secondly, it is interesting to note that the above system of equation differs from the one of Einstein-Cartan-Sciama-Kibble theory. Indeed, it has more degrees-of-freedom once the field equation \eqref{eq:mov2_coord} is not an algebraic equation for the spin connection and shows that, for a strong curvature regime, torsion generally behaves as a propagating field.

\section{Cosmology}
\label{sec:Cosmo}
The present Standard Cosmological Model (SCM) is a complex set of models that, combined together, gives us a coherent description of the evolution of the universe. Despite the delicate open issues, such as the nature of dark energy and dark matter, the SCM is a consistent framework that accounts for all present observations. The core of the SCM is the assumption that the universe is homogeneous and isotropic over scales larger than $ \SI{200}{\mega\parsec}$. The current CMB data seems to validate the isotropy of the universe around us but, since astronomical observations give access only to the past null-light cone, the homogeneity remains only as a profitable assumption.

The effective theory of gravity briefly described in the last section provides us a rich cosmological scenario. As will be shown, there are three distinctive regimes which, if properly connected, could outline the evolution of the universe from energies close to the Planck energy up to the $\si{\mega\electronvolt}$ scale where the SCM nucleosynthesis took place.

At the present analysis, we will use the term cosmological model for the evolution of a homogeneous and isotropic metric according to the field equations \eqref{eq:eq_mov1} and \eqref{eq:eq_mov2}. The matter content will be assumed to have negligible spin contribution and be represented as non-interacting fluids. In addition, in order to have a Riemannian spacetime, we will disregard torsion effects, i.e. the torsion 2-form will be assumed everywhere zero, $T^\mathfrak{a}(x)=0$. To make that explicit in the equations, all torsionless geometrical quantities will be denoted with the symbol $\circ{}$ upon them.

The energy scale of our cosmological model has a wide range of validity. Thus, we need to account for the QFT contribution to the cosmological constant. The net effect of this contribution is a change only in the cosmological constant term $\Lambda\rightarrow\tilde{\Lambda}=\Lambda+\Lambda_{\mathrm{QFT}}$ leaving intact the factor $1/\Lambda$ in the curvature square term (see last section and \cite{Sobreiro:2011hb,Assimos:2013eua}). Thus, our cosmological model will be based on the action
\begin{equation}\label{cosmol_action}
S=\int 
\left(\epsilon\mathcal{L}_{\mathrm{RR}}+\epsilon\mathcal{L}_{\mathrm{CC}}+\mathcal{L}_{\mathrm{EH}}+\mathcal{L}_{\mathrm{m}}\right)\;,
\end{equation}
where
\begin{IEEEeqnarray}{rCl}
\mathcal{L}_{\mathrm{RR}}&=&\frac{3}{32\pi G \Lambda}\tensor{\mathring{R}}{^{\mathfrak{a}}_{\mathfrak{b}}}\star\tensor{\mathring{R}}{_{\mathfrak{a}}^{\mathfrak{b}}}=\frac{3}{64\pi G\Lambda}\tensor{\mathring{R}}{_{\alpha\beta\mu\nu}}\tensor{\mathring{R}}{^{\alpha\beta\mu\nu}}\sqrt{-g}d^4x\;,\label{cosmol_actionRR}\\
\mathcal{L}_{\mathrm{CC}}&=&\frac{{\tilde\Lambda}}{192\pi G}\varepsilon_\mathfrak{abcd}e^\mathfrak{a}e^\mathfrak{b}e^\mathfrak{c}e^\mathfrak{d}=\frac{\tilde{\Lambda}}{8\pi G}\sqrt{-g}d^4x\;, \label{cosmol_actionCC}\\
\mathcal{L}_{\mathrm{EH}}&=&\frac{-1}{32\pi G}\varepsilon_\mathfrak{abcd}{R}^\mathfrak{ab}e^\mathfrak{c}e^\mathfrak{d}=\frac{-1}{16\pi G}\mathring{R}\sqrt{-g}d^4x\;,\label{cosmol_actionEH}
\end{IEEEeqnarray}
$\epsilon=\pm 1$, $\tensor{\mathring{R}}{^\alpha_{\beta\mu\nu}}$ is the Riemann curvature tensor and $\mathring{R}$ the Ricci scalar. As mentioned before, $\mathcal{L}_{\mathrm{EH}}$ and $\mathcal{L}_{\mathrm{CC}}$ denote, respectively, the Einstein-Hilbert and the cosmological constant Lagrangians and both of them are already present in General Relativity (GR). The Lagrangian $\mathcal{L}_{\mathrm{m}}$ sets the matter content that, as already said, will be considered as a combination of non-interacting fluids with negligible net spin. The extra lagrangian $\mathcal{L}_{\mathrm{RR}}$ deviates our dynamics from GR's by introducing the Kretschmann scalar $\tensor{\mathring{R}}{_{\alpha\beta\mu\nu}}\tensor{\mathring{R}}{^{\alpha\beta\mu\nu}}$ as a curvature square correction - even though it is suppressed by a $\Lambda$ denominator.

Consider a generic fluid with four-velocity field given by $v^\mu$. Its energy-momentum tensor can always be decomposed as
\begin{equation}\label{tmunugeneric}
\tensor{\mathcal{T}}{^{\mu\nu}}=\rho v^\mu v^\nu+p\tensor{h}{^{\mu\nu}}+v^{\mu}q^{\nu}+v^{\nu}q^{\mu}+\pi^{\mu\nu}\;,
\end{equation}
where the thermodynamic quantities $\rho$, $p$, $q^\mu$ and $\pi^{\mu\nu}$ are, respectively, the energy density, the pressure, the heat flux vector and the anisotropic pressure tensor. The $\tensor{h}{^{\mu\nu}}$ tensor is the projector defined as $\tensor{h}{^{\mu}}_{\nu}\equiv v^\mu v_\nu+\tensor{\delta}{^\mu}_\nu$.

In the presence of matter and for vanishing torsion, the field equations \eqref{eq:mov1_coord} and \eqref{eq:mov2_coord} become
\begin{IEEEeqnarray}{rCl}
\frac{3\epsilon}{8\Lambda}\tensor{\mathring{\Upsilon}}{^\mu_\nu}+\tensor{\mathring{G}}{^\mu_\nu}+\epsilon\tilde\Lambda\tensor{\delta}{^\mu_\nu}&=&\chi\,\tensor{\mathcal{T}}{^\mu}_\nu\;, \label{field1}\\
\mathring D_\alpha\Big(\Lambda^{-1}\ \tensor{\mathring{R}}{^\alpha_{\beta\mu\nu}}\Big)&=&0\;, \label{field2} 
\end{IEEEeqnarray}
where we introduced the gravitational coupling constant $\chi\equiv 8\pi G$ and $\tensor{\mathring{G}}{^\mu_\nu}$ is the symmetric Einstein tensor. Equation (\ref{field2}) can still be further simplified by using the Bianchi identities. In an arbitrary Riemannian geometry the Bianchi identities guarantee that $2\mathring{D}_\alpha \tensor{\mathring{R}}{^\alpha_{\beta}}=\mathring{D}_{\beta}\mathring{R}$. Thus, the second equation above can be recasted as
\begin{equation}
\frac{1}{\Lambda}\partial_\mu \mathring{R}=\frac{1}{\Lambda}\tensor{\mathring{R}}{_\mu^\alpha} \, \mathring{D}_\alpha \ln \Lambda\;.
\end{equation}

To consider a homogeneous and isotropic metric means that there is a special foliation where each spatial section is maximally symmetric. Therefore, the metric must be of a Friedmann-Lema\^itre-Robertson-Walker (FLRW) type and, hence, the interval can be written as 
\begin{equation}\label{flrw metric}
ds^2=-dt^2+a^2(t)\left(\frac{dr^2}{1-k r^2}+r^2\mathrm{d}\Omega^2\right)\;,
\end{equation}
where $a(t)$ is the scale factor, $\mathrm{d}\Omega^2$ is the solid angle and the constant $k \in \{-1,0,1\}$ defines the curvature of the spatial sections. 

In a geometric theory of gravity, isometries must come accompanied by symmetries of the energy-momentum tensor. In the particular case of a FLRW metric, the matter field acting as a source field must have an energy-momentum of a perfect fluid, i.e. equation \eqref{tmunugeneric} simplifies to
\begin{equation}\label{perffluidenergymomentum}
\tensor{\mathcal{T}}{^\mu_\nu}=(\rho+p)v^\mu v_\nu+p\tensor{\delta}{^\mu_\nu}\;.
\end{equation}

In the preferred coordinate system in which the interval takes the form (\ref{flrw metric}), the field equations reduce to a time evolution of the scale factor combined with homogeneous and isotropic thermodynamic quantities. It is useful to define two variables encoding time derivatives of the scale factor as
\begin{equation}\label{defs}
l\equiv \frac{\ddot{a}}{a} \;, \quad h\equiv \left(\frac{\dot{a}}{a}\right)^2+\frac{k}{a^2}\;.
\end{equation}
In terms of these variables, in a FLRW spacetime, the non-zero components of the relevant geometrical objects are
\begin{equation}
\tensor{\mathring{R}}{^i_j}=(l+2h)\tensor{\delta}{^i_j}\;, \;\; \tensor{\mathring{\Upsilon}}{^i_j}=4(2l^2+h^2)\tensor{\delta}{^i_j}\;,\;\;\tensor{\mathring R}{^0_0}=3l\;, \;\; \tensor{\mathring\Upsilon}{^0_0}=12h^2\;, \;\; \mathring R=6(l+h)\;.
\end{equation}
Defining two energy densities associated with the two cosmological constants as $\rhol\equiv \chi^{-1}\Lambda$ and $\rholl \equiv \chi^{-1}\tilde{\Lambda}$, the field equations read
\begin{IEEEeqnarray}{rCl}
\frac{3\epsilon}{2\chi\rhol}h^2-h+\frac{\chi}{3}(\rho+\epsilon\rholl)&=&0\;, \label{fields00}\\
\frac{3\epsilon}{2\chi\rhol}l^2-l-\frac{\chi}{6}(\rho+3p-2\epsilon\rholl)&=&0\;, \label{fieldsij}\\
\Lambda^{-1}\left[ \partial_t (l+h)- l \, \partial_t \ln\Lambda\right] &=&0\;.\label{fieldsextra}
\end{IEEEeqnarray}
This is the system of equations governing the evolution of this cosmological scenario. It must be mentioned that the identification \eqref{eq:newton-cosmol} induces an energy-dependent running on $\Lambda$ at a semi-classical level - this will be discussed in subsection \ref{subsec:deepUVsector}. In addition, we clearly have distinct contributions from each of the terms in the above system of equations. In fact, we can distinguish three main sectors of our model, namely, the infrared (IR), the ultraviolet (UV) and deep ultraviolet (deep UV).

The term ``ultraviolet'' is used in this context to explicit the fact that we are referring to a really high energy regime. In a geometric theory of gravity, this can be translated as saying that spacetime has a large curvature. However, both statements only make sense if the curvature (or energy) is being compared with some other equivalent quantity. This quantity would be the characteristic energy scale of the model, given by $\Lambda$ or, equivalently, by $\rhol$. For instance, we can compare the linear terms $h$ and $l$ with $\rhol$. If $h$ and $l\gg\rhol$, then the quadratic terms in \eqref{fields00} and \eqref{fieldsij} are comparable with the linear terms themselves and the full equations must be taken into account. In other words, the ``UV qualifier'' in this gravity context means that the UV correction terms cannot be neglected.

 In the following, we will analyze each of the three sectors going from low energy to very high energy scales.
 
\subsection{IR sector: connection to the $\Lambda$CDM model}

The IR sector is characterized by a low energy regime for gravity, i.e. relatively low spacetime curvature. In fact, if $h$ and $l\ll\rhol$, then the $\Lambda$-suppressed terms in \eqref{fields00} and \eqref{fieldsij} become negligible. This also means, as discussed in the end of section \ref{sec:Inducedgrav}, that the Cartan-like equation \eqref{field2} or, equivalently, \eqref{fieldsextra} can be disregarded completely. Effectively, one can reach the IR regime by taking the limit ${\Lambda}\rightarrow +\infty$. In this case, the dynamics is given by
\begin{IEEEeqnarray}{rCl}
h&=&\frac{\chi}{3}(\rho+\epsilon\rholl)\;, \label{fields00IR}\\
l&=&-\frac{\chi}{6}(\rho+3p-2\epsilon\rholl)\;, \label{fieldsijIR}
\end{IEEEeqnarray}
which are exactly the Einstein field equations with a cosmological constant. Note, however, that the sign of the cosmological term is determined by the parameter $\epsilon$. In particular, only if $\epsilon=1$ their solutions will contemplate the late time expansion of the $\Lambda$CMD model.

It is important to mention that the phase transition originating the present induced theory of gravity is expected to happen at energy scales of the order of $\SI{e16}{\tera\electronvolt}$. Therefore, the IR regime should be valid much before the primordial nucleosynthesis that occurred at \si{\mega\electronvolt}. Clearly, the IR cosmological sector mimics the $\Lambda$CDM model with an effectively tiny cosmological constant given by $\tilde{\Lambda}$.

\subsection{UV sector: high curvature and constant parameters regime}\label{subsec:UVsector}

Fairly below Planck energy any running effects in the gravitational parameters can still be safely neglected. In spite of this, we can also assume a high curvature situation. This scenario will be called the UV sector of this gravity theory and can be characterized by a finite and fixed $\Lambda$. The cosmological dynamics is given by \eqref{fields00}, \eqref{fieldsij} and
\begin{equation}
\partial_t (l+h)=0\;.\label{fieldsextraUV}
\end{equation}

In what follows we will outline separately the evolution of a vacuum and a matter filled spacetime.

\subsubsection{Vacuum case}\label{dS phase}\label{subsec:UVsector:vacuum}

In a vacuum universe, equations \eqref{fields00} and \eqref{fieldsij} resume to
\begin{IEEEeqnarray}{rCl}
\frac{3\epsilon}{2\Lambda}h^2-h+\frac{\epsilon \tilde{\Lambda}}{3}&=&0\;, \label{fields00UVvac}\\
\frac{3\epsilon}{2\Lambda}l^2-l+\frac{\epsilon \tilde{\Lambda}}{3}&=&0\;. \label{fieldsijUVvac}
\end{IEEEeqnarray}
These are algebraic relations for $h$ and $l$, respectively. In fact, both equations have the same structure showing that $h$ and $l$ share the same spectrum. It is straightforward to identify the roots as 
\begin{equation}\label{h-cte}
\lds\equiv\frac{\epsilon\Lambda}{3}\left(1\pm\sqrt{1-2\frac{\tilde{\Lambda}}{\Lambda}}\right). \\
\end{equation}
Since both $h$ and $l$ are constants, equation (\ref{fieldsextraUV}) is automatically satisfied. The Ricci scalar reads $\mathring R=6(l+h)$ which for $l=h$ gives $\mathring R=12 \lds$ and for $l\neq h$ gives $\mathring R=4\epsilon\Lambda$. The evolution of the scale factor can automatically be integrated from the equation $h=\lds$. The solutions are
\begin{equation}\label{solauv-vac}
a(t)=\frac{1}{\lds}\begin{cases}
\cosh \left(\sqrt{\slds} t \right) \quad ; \quad k=-1\;,  \\
\; \exp\left(\sqrt{\slds} t\right) \quad ;  \quad k=0\;,  \\
\sinh \left(\sqrt{\slds} t\right) \quad ; \quad k=1\;.
\end{cases}
\end{equation}
If $\epsilon=1$ one can immediately recognize these as being three different foliations of a de Sitter universe with an effective cosmological constant given by $\lds$. However, if $\epsilon=-1$, then $\lds$ is negative and the hyperbolic functions are, in fact, regular trigonometric functions. This latter case means that \eqref{solauv-vac} represents different foliations of an anti-de Sitter universe.

The deceleration parameter
\begin{equation}
q\equiv-\frac{\ddot{a}a}{\dot{a}^2}=\frac{-1}{1-k/(a^2\Lambda_{\text{dS}\pm})} \;,
\end{equation}
is always negative, manifesting the accelerated expanding phase. The effective cosmological constant $\Lambda_{\text{dS}\pm}$ depends on both $\tilde\Lambda$ and $\Lambda$. In the limit $\tilde\Lambda/\Lambda \ll 1$, we can expand it up to first order to obtain
\begin{equation}\label{soluv-expand}
\lds \approx \frac{\epsilon\Lambda}{3}\left[1\pm\left(1-\frac{\tilde\Lambda}{\Lambda}\right)\right].
\end{equation}
The root $\ldsplus$ is then approximately given by $\epsilon2\Lambda/3$ while the root $\ldsminus$ is approximately given by $\epsilon\tilde\Lambda/3$. Given the enormous value of $\Lambda$, if $\epsilon=1$ then the first root represents an universe with a violent de Sitter phase. Therefore, it may be associated with an inflationary expansion. On the other hand, the second root, also if $\epsilon=1$, would correspond to a smoothly accelerating phase, similar to the late time expansion in the $\Lambda$CDM model.

It is very important to mention the remarkable resemblance of these solutions with the trace-anomaly induced inflation (Starobinsky model), see e.g. \cite{Fabris:2000mq,Pelinson:2002ef,Koksma:2008jn} and references therein. Nevertheless, these two models have very different theoretical backgrounds and hence should be considered as alternatives to one another. More specifically, to trigger inflation, the trace-anomaly induced inflation makes use of loop-corrections coming from quantum matter fields conformally coupled to the classical FLRW geometry. The SO(m,n) gauge induced cosmological model has a de Sitter phase because of the presence of the Kretschmann scalar term \eqref{cosmol_actionRR} in the induced gravity action \eqref{cosmol_action}. This quadratic term in the ``field strength'' is an inheritance from the underlying Yang-Mills action. It does not need the presence of matter fields nor it makes use of conformal invariance.

\subsubsection{Matter case} \label{subsec:UVsector:matter}

A generic perfect fluid has an energy-momentum tensor given by eq. \eqref{perffluidenergymomentum}. To complete its specification, one has to provide an equation of state which, in cosmology, is generally a functional dependence of the pressure in terms of the energy density, i.e. $p=p(\rho)$. Fluids that obey this equation of state are known as barotropic fluids. In GR the fluid equation of state is of major importance to completely specific the dynamical system. We are about to see that this is not the case for the UV sector of our model.

The dynamics of the matter filled spacetime is given by the system \eqref{fields00}, \eqref{fieldsij}, and \eqref{fieldsextraUV}. In principle, we have a system of three equations and three variables, namely, $a(t)$, $\rho(t)$ and $p(t)$ that will determine the evolution of spacetime and matter. Taking into account the thermodynamic equation of state would only make this system overdetermined. Indeed, equations \eqref{fields00}, \eqref{fieldsij} and \eqref{fieldsextraUV} are solvable without the need of an equation of state. A possible way to reconcile this situation with a thermodynamic description of matter is to interpret the UV sector as a regime where the gravitational field does not distinguish the nature of the matter fields. In other words, in the UV sector any perfect fluid gravitates in the same manner.

To solve the above system of equation, we first focus on \eqref{fieldsextraUV}. This equation states that the Ricci scalar, $\mathring R=6(l+h)$, has to be a constant which we write as $r_0$. Thus, we can substitute $l=r_0/6-h$ in \eqref{fieldsij} to obtain a new equation for the variable $h$, which can be combined with \eqref{fields00} to provide
\begin{equation}
h=\frac{\chi^2\rhol}{4\chi\rhol-r_0}(\rho+p)+\frac{r_0}{12}\;. \label{fieldsUVh}
\end{equation}
Note, however, that from the definition of $h$ itself - see \eqref{defs} -  we have $\dot{h}=2\dot{a}{a}^{-1}(l-h)$ and hence
\begin{equation}
\dot{h}=\frac13\frac{\dot{a}}{a}(r_0-12h)\quad  \Rightarrow \quad  h=\frac{\xi_0}{a^4}+\frac{r_0}{12}\;,\label{fieldsUVh2}
\end{equation}
where $\xi_0$ is a constant of integration. The above equation can be further integrated to obtain the time evolution of the scale factor.
In order to do this, we recast \eqref{fieldsUVh2} by writing $a^2(t)=x(t)$ such that
\begin{equation}\label{eqdiff1}
\dot{x}^2-\frac{r_0}{3}x^2+4 k x=4 \xi_0\;.
\end{equation}
The solutions of this differential equation are branched in basically three situations. When $r_0 \neq 0$ we have
\begin{equation}\label{gensol}
x(t)=x_0e^{\pm \alpha t}+\frac{9k^2-3r_0\xi_0}{R^2_0x_0}e^{\mp\alpha t}+\frac{6k}{r_0}\;,
\end{equation}
where $\alpha=\sqrt{r_0/3}$ and $x_0>0$ is a constant of integration. Of course, the characteristic of these solutions depends on the interplay among the constants $k$, $r_0$ and $\xi_0$. In particular, it can describe a bounce if the second term has the same sign as the first one, namely, if $r_0\xi_0<3k^2$.

For a null Ricci scalar $r_0=0$ but nonzero spatial section, the scale factor evolves accordingly to
\begin{equation}\label{solxricci0}
a(t)=\sqrt{{\xi_0}{k}-k\left(t\pm \sqrt{|\xi_0|}\right)^2}\;,
\end{equation}
where we chose the origin of time to correspond to the (classical) singularity. The $\pm$ sign within the square term does not change qualitatively the evolution. For $k=1$ we have a Big Bang\textbackslash Big Crunch solution with an initial and a final singularity. The scale factor has maximal range of $\Delta t=2\sqrt{\xi_0}$ where $a_{max}=\sqrt{\xi_0}$. For $k=-1$ we have two disjoint branches: either an expanding universe with initial singularity or a collapsing universe with a future singularity. For instance, for the plus sign the initial singularity is located at $t=0$, while the final singularity in the collapsing phase is at $t=-2\sqrt{\xi_0}$.

\begin{figure}[htb]
	\centering
		\includegraphics[width=.8\textwidth]{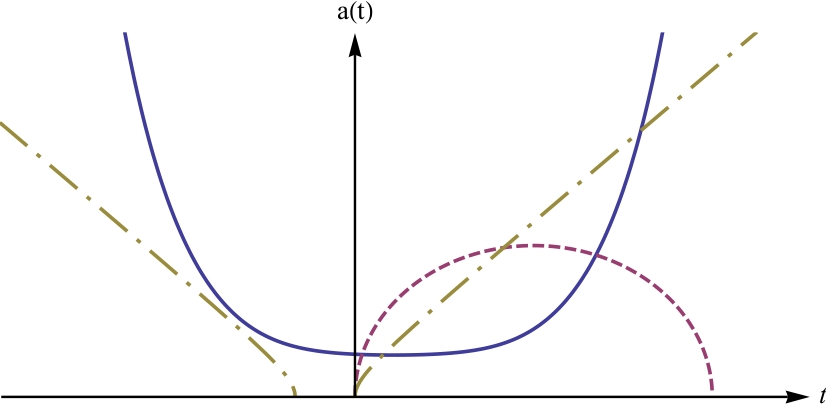}
		\caption{Three possible qualitatively distinct behavior of the scale factor. The solid line describes a non-singular bounce dynamics. The dashed line depict a typical Big Bang/Big Crunch model starting from and ending on a singularity. The dot-dashed lines represent two other possible singular models either starting from a singularity and expanding forever or an universe that contracts from infinity and ends on a singularity.}\label{fig3}
\end{figure}

Finally, for a flat spatial section, we have that the constant $\xi_0$ must be positive-definite and the solution is given by
\begin{equation}\label{solxricci0k0}
a(t)=\begin{cases}
\sqrt{-2\sqrt{\xi_0}t}& \text{if $t<0$}\;,\\
\sqrt{+2\sqrt{\xi_0}t}& \text{if $t>0$}\;.
\end{cases}
\end{equation}
where the constant of integration was chosen to locate the (classical) singularity at $t=0$. Once more we have a collapsing phase for $t<0$ that reaches the singularity and an expanding phase originating at an initial singularity at $t=0$. For a qualitative view of all these solutions, see fig. \ref{fig3}.

\begin{figure}[htb]
	\centering
		\includegraphics[width=.8\textwidth]{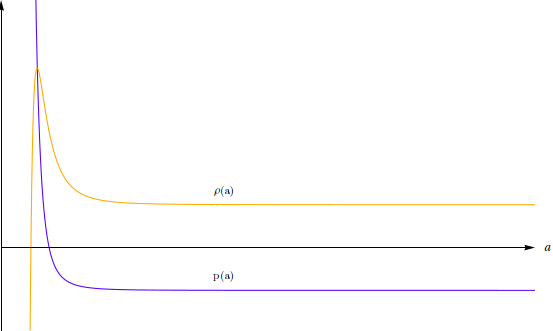}
		\caption{Plot of pressure and energy density as a function of the scale factor. Note that for large values of scale factor both become constants with $p=-\rho$. In particular, the values of the parameters are $\xi_0=0.001$, $\chi \rhol=4.1$, $\chi \rholl=1.12$ and $r_0=1.66$.}\label{fig1}
\end{figure}

\begin{figure}[htb]
	\centering
		\includegraphics[width=.8\textwidth]{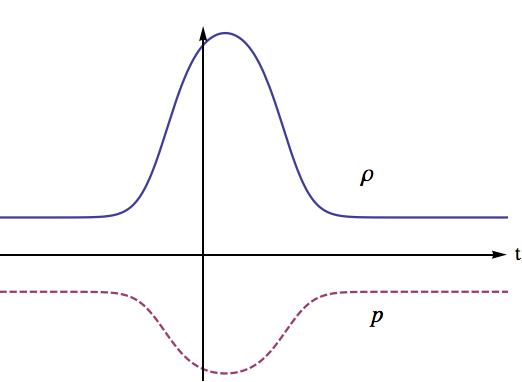}
		\caption{Time evolution of the pressure and energy density through the bounce. The values of the parameters are $\xi_0=1.8$, $\chi \rhol=2.0$, $\chi \rholl=0.32$ and $r_0=1.66$.}\label{fig2}
\end{figure}

The dynamics of the energy density and pressure can be found using the other two equations. Eq. \eqref{fieldsUVh2} can be combined with equations \eqref{fields00} and \eqref{fieldsUVh} resulting in the evolution of the two thermodynamic quantities
\begin{IEEEeqnarray}{rCl}
\rho&=&\frac{3\xi_0(4\chi \rhol- r_0)}{4\chi^2 \rhol}a^{-4}-\frac{9{\xi_0}^2}{2\chi^2\rhol}a^{-8}-\left(\rholl-\frac{r_0}{4\chi}+\frac{r_0^2}{32\chi^2\rhol} \right)\;, \label{exo_rho} \\
p&=&-\rho+\left(\frac{4\chi\rhol-r_0}{\chi^2 \rhol}\right)\xi_0a^{-4}\;.\label{exo_p}
\end{IEEEeqnarray}
The particular case $\xi_0=0$ freezes the value of the energy density and the pressure becomes $p=-\rho$. This case reflects a simple redefinition of $\rholl$ and hence we will assume $\xi_0\neq0$. The solutions \eqref{exo_rho} and \eqref{exo_p}, plotted in fig. \ref{fig1}, show the universal behavior of the pressure and energy density, independently of the equation of state of the fluid. Of course, we can use the solutions for the scale factor to obtain $\rho(t)$ and $p(t)$. For instance, see fig. \ref{fig2}. On the other hand, if we had assumed a barotropic equation of state of the fluid, say $p=w \rho$ with constant $w$, we would have found that equations \eqref{fieldsUVh} and \eqref{fieldsUVh2} necessarily implies $w=-1$. Thus, this model in its UV sector does not accept a barotropic equation of state describing matter. In fact, the only consistent barotropic equation of state in the UV sector is for the observational cosmological constant, $p_{\tilde{\Lambda}}=-\rholl$.

As we can observe, Figure~\ref{fig1} indicates that extrapolating equation \eqref{exo_rho} for very small values of the scale factor the energy density goes to minus infinity. Notwithstanding, equations \eqref{exo_rho} and \eqref{exo_p} are valid only for the UV sector. For higher energies (in the deep UV sector), these equations are no longer valid, in other words, we assume that the UV sector is valid for a range of the scale factor bigger than a critical value beyond which the energy density is already positive. This is consistent with the fact that the model is only valid below Planck scale. Hopefully, the negative energy sector can be fine tuned to exist only beyond Planck scale, where the theory is entirely different from standard gravity theories. However, we are aware that this is not the full resolution because there is also the risk of the energy density to become negative again in the future, i.e. for large values of the scale factor. This issue can be avoided by imposing limits on the two integration constants. The global maximum of the energy density happens at 
\begin{equation}\label{a-critical}
a_c=\left(\frac{12\xi_0}{4\chi\rho_\Lambda-r_0}\right)^\frac{1}{4}~.
\end{equation}
and hence $r_0<4\chi\rho_\Lambda$. The asymptotic energy density limit is
\begin{equation}
\rho_\infty=\lim_{a\rightarrow \infty} \rho(a) = -\frac{r_0^2}{32\chi^2\rho_\Lambda}+\frac{r_0}{4\chi}-\tilde{\rho}~.
\end{equation}
Therefore the condition 
\begin{equation}
1-\sqrt{1-2\frac{\rho_{\tilde{\Lambda}}}{\rho_\Lambda}}<\frac{r_0}{4\chi\rho_\Lambda}<1
\end{equation}
guarantees that the UV regime is free of negative energy density. As a note, we stress that in general $\rho_{\tilde{\Lambda}}\ll \rho_\Lambda$, so we can approximate the above condition by $4\chi \rho_{\tilde{\Lambda}} <r_0< 4\chi\rho_\Lambda$.

There is another important point about the values of the scale factor that would leave the energy density with negative values. The equation \eqref{exo_rho} has one possible real root,
\begin{equation}
a_{null}=\left\{\frac{12\xi_0}{(4\chi\rho_\Lambda-r_0)}\left[ 1+\sqrt{1+2\rho_\Lambda\left(\frac{4\chi}{4\chi\rho_\Lambda-r_0}\right)^2\rho_\infty} \right]^{-1}\right\}^\frac{1}{4}~,
\end{equation}
which clearly permit us to control the negative region with the $\xi_0$ since $a_{null}\propto \sqrt[4]{\xi_0}$. Therefore, as we can observe in the figure \eqref{fig1}, the negative region is meaningless once we can achieve the smallest values for $\xi_0$. Of course, it is also confirmed with the critical value for the scale factor in equation \eqref{a-critical}. Thus, $a_{null}$ as well as $a_c$ stand close to the origin when $\xi_0$ takes very small values.

In a FLRW universe, the conservation of the energy-momentum tensor of a barotropic fluid provides the energy density in terms of the scale factor. In particular, if $p=w \rho$ then $\rho\propto a^{-3(1+w)}$. Hence, it is clear that the energy density and pressure given by \eqref{exo_rho} and \eqref{exo_p} must be associated with a non-conservation of the matter energy-momentum tensor. Indeed, taking the time derivative of equation \eqref{fields00} and using equations \eqref{fieldsij} and \eqref{fieldsUVh2} we arrive at
\begin{equation}
\dot{\rho}+3\frac{\dot{a}}{a}(\rho+p)+\frac{\dot{\rho}^2}{8(\dot{a}/a)(\rho+\rholl-\rhol/2)}=0 \;,
\end{equation}
confirming the non-conservation of the matter energy-momentum tensor. The first two terms are the usual components while the last non-linear term comes from the geometrical UV corrections.

\subsection{Deep UV sector: running in $\rhol$}\label{subsec:deepUVsector}

In Section~\ref{sec:Inducedgrav}, we briefly described the emergency of a geometrodynamical theory of gravity from a pure Yang-Mills theory. The energy scale of this phase transition turned out to be in the same order of magnitude of the Planck energy. Therefore, the period of this transition is compatible with the end of what is known in cosmology as the Planck Era. In this sense, we expect that such classical theory of gravity could be valid up to very high energies.

One of the most important hypothesis used in building this induced geometrodynamical theory of gravity is expressed in equation \eqref{eq:newton-cosmol}, which identifies uniquely the gravitational parameters, $i.e.$, Newton's constant $G$ and the (renormalized) gravitational cosmological constant $\Lambda$, with the quantum ones, $i.e.$, the Gribov mass parameter $\gamma^2$ and the coupling parameter $\kappa^2$. After the phase transition, it is possible to fix one of the gravitational parameter, for instance, Newton's constant, to estimate the other, the (renormalized) cosmological constant -- see the details of such estimates in \cite{Sobreiro:2016fks,Assimos:2013eua}. Nonetheless, during the phase transition, which happens around Planck scale, $\Lambda$ or $G$ may vary as a consequence of the running of the renormalized quantum parameters and the identification \eqref{eq:newton-cosmol}. Here we assumed a varying $\Lambda$ and a fixed $G$ during this period, which we are referring to as the deep ultraviolet (deep UV) sector.

The dynamics of the deep UV sector is governed by the full system of equations as displayed in \eqref{fields00}, \eqref{fieldsij} and \eqref{fieldsextra}. To solve this system, it is necessary to have the knowledge of the specify time dependence of $\Lambda$. Unfortunately, this dependence is closely tied to the behavior of $\gamma^2$ in the non-perturbative regime of the original Yang-Mills theory - where the identification \eqref{eq:newton-cosmol} took place. As perturbative techniques are no longer valid, we lack a method to evaluate $\Lambda(t)$ and thus we have no firm ground to go beyond our remarks. We will leave this sector for future investigations.

Fortunately, the running of $\Lambda$ saturate already at ultra high energies, which is consistent with the other regimes described above. In fact, the semi-perturbative analysis in \cite{Assimos:2013eua,Sobreiro:2016fks} indicate that the deep UV epoch only lasts for $\approx 10^{-44}s$, which is right below Planck time.

\section{Conclusions}
\label{sec:Concl}

The present standard cosmological model is heavily based upon the FLRW metric. In this work we addressed the FLRW-like models in the effective gravity scenario discussed in section \ref{sec:Inducedgrav}. The induced gravity theory developed can be seen as a non-Riemannian generalization of GR with UV completion terms. As a first approach, we assumed a Riemannian spacetime which can be accomplished by annulling all torsion's degrees of freedom. Notwithstanding, the UV correction terms have shown to be sufficient to bring new features to the dynamics of the model.

In section \ref{sec:Cosmo} we showed that our model displays three distinct regimes. These dynamical regimes can be identified by comparing the curvature invariants with the bare cosmological constant $\Lambda$. In the IR sector, which is defined as a low curvature regime with negligible UV contributions, the theory mimics GR with an effective cosmological constant $\tilde{\Lambda}$. For a homogeneous and isotropic universe we naturally recover the $\Lambda$CDM model, hence smoothly connecting with the standard cosmological model. In order to secure all present early universe observation we need to connect our model with the $\Lambda$CDM at least before nucleosynthesis. Indeed, as long as the bare cosmological constant suppressing the UV corrections is really high, $\Lambda^{1/2} \SI{\sim e16}{\tera\electronvolt}$, we expect that the IR regime is reached much before the primordial nucleosynthesis which happened at \si{\mega\electronvolt}. Nevertheless, we should emphasize that this feature is only true for the $SO(5)$ induced gravity analyzed in the present work. Different $SO(m,n)$ induced gravity models might be in contradiction with the standard cosmological model.

The UV sector is characterized by a high curvature regime which is of the same order of magnitude of $\Lambda$. In this regime, the energy scale is comparable with customary inflationary phase. We analyzed separately a vacuum universe and a perfect fluid permeated spacetime. In the former, equations \eqref{fields00UVvac} and \eqref{fieldsijUVvac} were solved and we obtained three solutions for the scale factor, depending on $\epsilon$ and the curvature of the spatial section as listed in \eqref{solauv-vac}. In order to obtain an expanding de Sitter phase one must consider only the $\epsilon=1$ solutions. This is the same condition that was required to associate the IR sector with the $\Lambda$CDM model. Hence, for the $SO(5)$ induced gravity, the cosmological model can have a de Sitter primordial phase consistently connected with a $\Lambda$CDM universe.

The induced gravity theory was developed in a first order formalism which increase the number of dynamical equations. For GR, the extra field equation associated with the spin connection's degrees of freedom is not properly a dynamical equation. In fact, it establishes the affine nature of the spacetime by requiring the connection to be identified with Christoffel symbols. In our case, we do have an extra dynamical equation that turns the set of cosmological equations into a determined system. As a consequence, one cannot introduce an equation of state for the perfect fluid. As we have argued in section \ref{subsec:UVsector:matter}, the proper way to interpret this result is to consider that, in the UV regime, gravity do not distinguish different fluids. All perfect fluids gravitate in the same manner. Among the possible solutions, there are singular solutions with single past or future singularities, Big Bang\textbackslash Big Crunch and also nonsingular solution with a single bounce that can be symmetric or antisymmetric.

The deep UV sector is also a high curvature regime. However, the main difference with respect to the UV sector is a possible running on $\Lambda$. This energy dependence comes from the running of the coupling parameter $\kappa^2$. The dynamic systems governing this period is enclosed in equations \eqref{fields00}-\eqref{fieldsextra}. The energy scale of the deep UV sector is in the order of Planck energy and, to adequately describe this regime one needs to specify the behavior of $\gamma^2$ in the non-perturbative sector of the original Yang-Mills theory. Unfortunately, at present moment, we lack a method to evaluate this dependence since perturbative techniques are no longer valid.

\section*{Acknowledgements}

The Conselho Nacional de Desenvolvimento Cient\'ifico e Tecnol\'ogico (CNPq - Brazil), the Coordena\c{c}\~ao de Aperfei\c{c}oamento de Pessoal de N\'ivel Superior (CAPES) and the Pr\'o-Reitoria de Pesquisa, P\'os-Gradua\c{c}\~ao e Inova\c{c}\~ao (PROPPI-UFF) are acknowledge for financial support.

\bibliographystyle{utphys2}
\bibliography{library}

\end{document}